\documentclass[preprint,showpacs,preprintnumbers,amsmath,amssymb]{revtex4}

\usepackage{graphicx}
\usepackage{epstopdf}
\usepackage{dcolumn}
\usepackage{bm}

\newcommand{\be}{\begin{equation}}
\newcommand{\ee}{\end{equation}}
\newcommand{\bea}{\begin{eqnarray}}
\newcommand{\eea}{\end{eqnarray}}

\newcommand{\Eq}[1]{Eq.\,(\ref{#1})}

\def\NHN{{\textrm{N-H}\cdots\textrm{N}}}
\def\OHO{{\textrm{O-H}\cdots\textrm{O}}}
\def\NHO{{\textrm{N-H}\cdots\textrm{O}}}
\def\NN{\textrm{N}\cdots\textrm{N}}
\def\NO{\textrm{N}\cdots\textrm{O}}

\def\cm{\textrm{cm}$^{-1}\,$}

\begin{document}

\title{QM/MM Simulation of the Hydrogen Bond Dynamics of an Adenine:Uracil Base Pair
in Solution. Geometric Correlations and Infrared Spectrum}

\author{Yun-an Yan}
\author{Oliver K\"uhn}
\email{oliver.kuehn@uni-rostock.de}
\affiliation{
Institut f\"{u}r Physik, Universit\"{a}t Rostock, D-18051 Rostock, Germany
}%

\date{\today}

\begin{abstract}
Hybrid QM(DFT)/MM molecular dynamics simulations have been carried out for the Watson-Crick base pair 
of 9-ethyl-8-phenyladenine and 1-cyclohexyluracil in deuterochloroform solution at room temperature. 
 Trajectories are analyzed putting special attention to the geometric 
correlations of the $\NHN$ and $\NHO$ hydrogen bonds in the base pair. Further, based on empirical correlations between 
the hydrogen bond bond length and the fundamental NH stretching frequency its fluctuations are obtained along the trajectory. Using the
time dependent frequencies the infrared lineshape is determined assuming the validity of a second order 
cumulant expansion. The deviations for the fundamental transition frequencies are calculated  to amount to less than 2\%  as compared with experiment. The width of the spectrum  for the $\NHN$ bond is in reasonable agreement with experiment while that for the $\NHO$ case is underestimated by the present model. Comparing the performance of different pseudopotentials
 it is found that the Troullier-Martins pseudopotential with a 70 Ry cut-off yields the best agreement.
\end{abstract}

\maketitle

\section{Introduction}
One of the primary examples for the importance of hydrogen bonding is its role 
in the structural selectivity of the pairing of nucleic acid bases in 
DNA \cite{jeffrey97}. Here double and triple hydrogen bonds (HBs) are formed 
whose theoretical characterization still poses a challenge as it combines many 
features of multidimensional condensed phase quantum dynamics.  Infrared (IR) 
spectra are particularly sensitive to HB formation. The broad lineshapes not only contain 
information on the intramolecular couplings, but the underlying 
fluctuation dynamics also reflects the interaction with the surrounding 
medium \cite{giese06:211,nibbering07:619}. The spectroscopic signatures of hydrogen bonding 
in heterodimers of various bases have been investigated in different phases. Isolated adenine:thymine 
pairs, for instance, were studied in gas phase using IR-UV double resonance 
experiments \cite{plutzer03:838}. This provided evidence that Watson-Crick 
pairing does not result in the most probable form under these conditions. 
Later, the dominant tautomer could be identified on the basis of quantum 
dynamical simulations of IR spectra including anharmonicity \cite{krishnan07:132}. 
Ultrafast nonlinear IR spectroscopy has been used to scrutinize the vibrational 
dynamics of individual base pairs in solution \cite{woutersen04:5381}. Here it 
was found, for instance, that the life time of the H-bonded NH stretching 
fundamental transition decreased by a factor of three as compared to the monomer 
case, thus demonstrating the effect of HB mediated anharmonic couplings.
The next step in complexity has been addressed by ultrafast two-color IR studies 
of the HBs in DNA oligomers. Heyne et al. showed that 
excitation in the fingerprint region provides, by virtue of the anharmonic 
couplings, a means to identify, for instance, the symmetric NH$_2$ stretching 
vibration of adenine:thymine pairs upon probing in the 3200 \cm 
range \cite{heyne08:7909}. In the linear IR spectrum this range is completely 
masked by the absorption of water molecules. Subsequently, the assignment has 
been confirmed by anisotropy measurements \cite{dwyer08:11194}.

Since first principles quantum dynamical simulation of condensed phase HB 
dynamics will remain elusive different approximate strategies are usually 
followed. In order to capture the electronic structure associated with H-bonding  
density functional theory (DFT) seems to be most appropriate as long as stacking 
interactions are of minor importance \cite{hobza99:3247}. By running classical 
trajectories with on-the-fly DFT forces the IR spectrum can be obtained from 
the Fourier transform of the dipole autocorrelation function. Of course, this 
completely neglects quantum effects in the nuclear motion. The latter are 
frequently accounted for by quantum correction factors \cite{ramirez04:3973}. 
Car-Parrinello molecular dynamics (CPMD) has proven to be particularly 
well-suited for simulating condensed phase dynamics including H-bonded systems. 
A pioneering contribution has been given by Silvestrelli et al. who determined 
the IR spectrum of liquid water \cite{silvestrelli97:478}. Subsequent 
applications to  aqueous solutions, e.g., of uracil have been summarized in 
Ref. \cite{gaigeot03:10344}. More recent studies focussed on strongly H-bonded 
system like the Zundel cation in the crystalline phase \cite{vener05:258}. 

In principle quantum effects for the fast proton stretching motion can be 
recovered by calculating its potential energy curve for selected configurations along the CPMD trajectory (snapshot potentials). 
Upon solving the stationary Schr\"odinger equation for these potentials one 
obtains a set of fundamental transition frequencies. The combination of CPMD 
and snapshot potentials for the proton stretching motion has been put forward 
in Ref. \cite{jezierska07:205101}. In this reference an intramolecular $\NHO$ HB 
in a Mannich base in gas and crystalline phase was considered. The IR spectrum 
has been calculated by convolution of the stick spectrum obtained along the 
trajectory with a Gaussian. This gave a band maximum below 2000 \cm, considerably  
red shifted  with respect to the estimated experimental value. Interestingly, 
the analogous simulation in CCl$_4$ solution gave a band whose first moment has 
been well above 2000 \cm, pointing out the sensitivity with respect to the 
environment \cite{jezierska07:5243}. In Ref.  \cite{jezierska08:839} the method 
was applied to an $N$-oxide which exhibits two rather strong nonequivalent 
$\OHO$ HBs in the unit cell. Here, two-dimensional snapshot potentials 
including the bending vibration of the HB had been considered as well. 
Finally, a different $N$-oxide in the crystalline phase with a rather strong 
HB was studied in Ref. \cite{stare08:1576}. In this case a 1000 \cm broad 
distribution of transition frequencies for the OH-stretching fundamental was 
found from one-dimensional snapshot potentials in reasonable agreement with experiment. 

Quantum Mechanics/Molecular Mechanics (QM/MM) hybrid methods have been developed 
to  cope with situations where only a part of a large system needs to be treated 
at the quantum mechanical level (for a review, see Ref. \cite{lin07:185}). Here the crucial point is the 
interface between the QM and MM parts and in the context of CPMD different 
schemes for the most delicate issue of the nonbonding electrostatic interactions 
have been developed \cite{laio02:6941,biswas05:164114}. A series of impressive 
applications to the IR spectroscopy of various forms of water clusters in a 
model of bacteriorhodopsin has been presented in 
Refs. \cite{rousseau04:4804,mathias07:6980,baer08:2703}.

Recently, we have applied the QM/MM approach to the determination of the IR 
lineshape of the uracil NH stretching vibration in a modified adenine:uracil 
(A:U) pair in deuterochloroform solution \cite{yan08:230}. In doing so we made 
use of an empirical NH-frequency-$\NN$-distance correlation which gave access 
to the fluctuating transition frequency along the trajectory (see also 
Ref. \cite{bratos04:197}). From the autocorrelation function of this  
transition frequency the lineshape could be determined. This procedure gave 
the band maximum at 3209 \cm and the full-width at half maximum (FWHM) of 39 \cm, 
both values being in reasonable agreement with experiment \cite{woutersen04:5381}. 

In the present contribution we will extend our previous study in several 
respects. First, the IR spectral features of  both intermolecular  HBs are 
investigated. This requires to establish a distance-frequency correlation for 
$\NHO$ HBs which is done on the basis of available crystal structure data. 
Alternatively, we explore the possibility to use snapshot potentials for 
obtaining this correlation curve. It will turn out that the snapshot frequencies 
underestimate the average transition frequency considerably and are of little 
use for the considered intermolecular HBs. In order to scrutinize this issue 
we report an investigation of the influence on the used 
pseudopotential (PP) and kinetic energy cut-off parameter in the DFT calculation. Besides the IR spectrum we 
will put emphasis on the geometric correlations across the double HBs which can be 
obtained along the trajectory. This concerns in particular the linearity, 
planarity, and the correlation between HB distance and proton position.

The paper is organized as follows: In the next Section II we start by defining the model system and the QM/MM protocol. Further,
the issues of geometric correlations and lineshape analysis are discussed. In Section III we first give results on geometric correlations before we come to the lineshape and its dependence on the PP used. The results are summarized in Section IV.
%
\section{Methods}
\subsection{QM/MM Protocol}
\begin{figure}[ht!]
\includegraphics[width=0.45\textwidth]{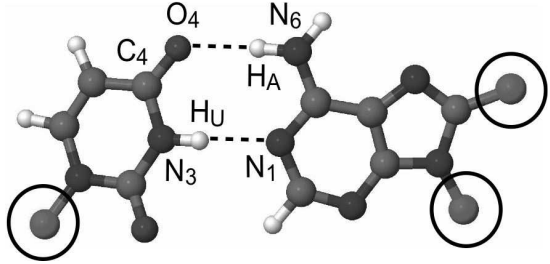}
\caption{QM part of the A:U base pair with the capping atoms indicated by circles. 
In the analysis below we will use the angles  $\alpha$ (N$_6$-O$_4$-C$_4$) 
and $\beta$ (C$_4$-N$_3$-N$_1$), the 	angle $\gamma$ between the two vectors 
$\overset{\longrightarrow}{\textrm{N}_3\textrm{N}_1}$ and     
$\overset{\longrightarrow}{\textrm{O}_4\textrm{N}_6}$ as well the dihedral 
angle $\phi$ defined by the atoms N$_{1}$, N$_{3}$, O$_{4}$, and N$_{6}$.}
\label{fig:structure}
\end{figure}

We will investigate the vibrational dynamics of 9-ethyl-8-phenyladenine (A) 
and 1-cyclohexyluracil (U)  solvated in CDCl$_3$ which has been studied 
experimentally by Woutersen and Cristalli \cite{woutersen04:5381}. To this end a hybrid QM/MM 
method is used as implemented in the Gromacs/CPMD interface \cite{biswas05:164114}. 
All atoms of the A:U pair except those belonging to the substituents are dealt 
with quantum mechanically by  the CPMD package \cite{cpmd}. The substituents themselves 
as well as the solvent are described by the OPLS all-atom force 
field \cite{jorgensen96:11225} as implemented in Gromacs \cite{gmx31}. H-atom 
capping is used to  saturate the dangling bonds at the QM/MM boundary 
(see Fig.~\ref{fig:structure}).
In the simulation a single  A:U pair is solvated in 100 $\textrm{CDCl}_3$ 
molecules within a box having dimensions 
$30.0~\textrm{\AA}\times23.5~\textrm{\AA}\times23.5~\textrm{\AA}$ 
(density is 0.1 M \cite{woutersen04:5381}). 
The QM part is placed in a 
$21.2~\textrm{\AA}\times15.9~\textrm{\AA}\times15.9~\textrm{\AA}$ box. 
The Becke exchange and Lee-Yang-Par correlation functional (BLYP) together 
with the plane wave basis set is used as implemented in the CPMD code. 
The MM molecular dynamics run is performed  at  298 K  using a time step 
of 2 fs. Note that in the current Gromacs/CPMD implementation it is 
actually a  Born-Oppenheimer simulation which is performed.
For the  initial configuration  the geometry of the optimized 
gas phase base pair replacing solvent atoms in the equilibrated simulation box 
has been used. First, a 1 ps equilibration run was performed using the 
Vanderbilt (VB)  ultrasoft PP  with a plane wave  cutoff of 30 Ry. Subsequently, 
trajectories were propagated up to 5.0 ps using different PPs and cut-off parameters, 
i.e., VB \cite{vanderbilt90:7892} with a cut-off of 
30 Ry and 40 Ry, Troullier-Martins (TM) PP  \cite{troullier91:1993} with 70 Ry and 100 Ry, 
and Goedecker (GO) PP  \cite{geodecker96:1703} with 100 Ry and 140 Ry. Except for the comparative studies the results presented are based on the TM PP using a cut-off of 70 Ry.
%
\begin{figure*}[ht!]
\includegraphics[width=\textwidth]{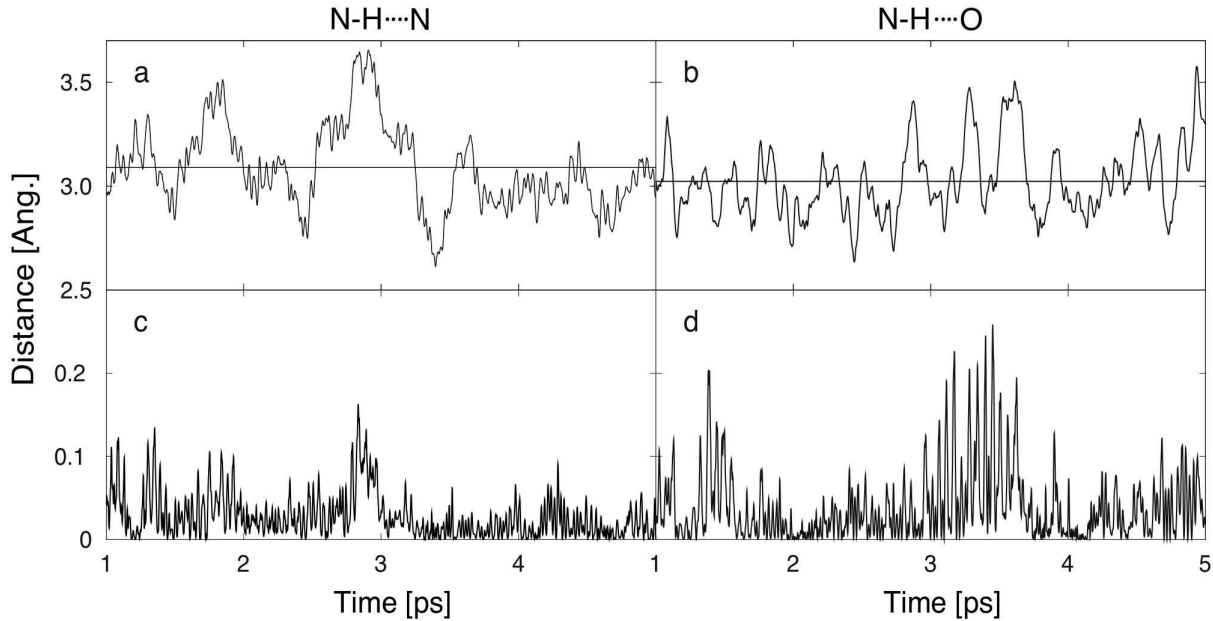}
\caption{Geometric parameters of the HBs along the trajectory:
(a) and (b) HB lengths of $\NHN$ and 
$\NHO$, respectively. The horizontal line indicates the 
time average. (c) and (d) out-of-line motion of the H atom  in the 
two HBs, measured by the difference
$L(\textrm{N-H}) + L(\textrm{N}\cdots\textrm{H}) - L(\textrm{N}\cdots\textrm{N})$
and
$L(\textrm{N-H}) + L(\textrm{O}\cdots\textrm{H}) - L(\textrm{N}\cdots\textrm{O})$,
respectively.
}
\label{fig:dist}
\end{figure*}
%

%
\subsection{Hydrogen Bond Geometry}
Hydrogen bonds A-H$\cdots$B display remarkable  correlations in their geometry. 
Limbach and coworkers promoted a model which gives a simple explanation of 
these correlation in terms of valence bond orders $p_i$ viewing the HB 
as composed of two diatomic units A-H and B-H with bond lengths $R_1$ 
and $R_2$, respectively \cite{limbach06:193}:
\begin{equation}
\label{eq:bondorder}
p_i=\exp\{-(R_i-R_i^{\rm eq})/b_i\}\, ,
\end{equation}
where $R_i^{\rm eq}$ is the equilibrium bond length of the hypothetical 
nonbonded diatomic and $b_i$ is a parameter describing the change of the 
bond valence upon bond stretching. If one assumes that the total bond order 
is unity, i.e. $p_1+p_2=1$, the two bond lengths cannot be changed 
independently. This can be alternatively expressed in terms of the deviation 
from the HB center $(R_1-R_2)/2$ and the HB length $R_1+R_2$.  
The correlation curve so obtained from numerous structural data expresses a fact 
well known from quantum chemical studies of potential energy surfaces, 
namely that the  HB is compressed upon H transfer while passing the 
transition state  \cite{giese06:211}. This correlation for a given type of 
HB - although expressed in Eq. (\ref{eq:bondorder}) by four parameters 
only - yields  a rather robust description of the HB geometries 
(see, e.g. the case of N-H$\cdots$N bonds in 
Refs. \cite{limbach06:193,pietrzak07:296}). It turns out, however, that 
in particular for short HBs the original formulation required some 
modification. In Ref. \cite{limbach04:115} Limbach et al. 
suggested an empirical correction as follows
\begin{equation}
\label{eq:bondordercor}
p_{1/2}^{\rm H}= p_{1/2} \mp c (p_1-p_2)(p_1p_2)^5 - d (p_1p_2)^2\, ,
\end{equation}
for which $p_{1}^{\rm H}+p_{2}^{\rm H}<1$ holds. Here, $c$ and $d$ are 
parameters to be determined by fitting geometric correlations to experimental data.

The change in the potential energy surface which is reflected in the above 
geometric correlation also manifests itself in a dependence of the AH stretching 
frequency on the HB length. In a seminal work Novak collected experimental 
data on different H-bonded crystals to establish A-B bond length-frequency 
correlation for  $\NHN$ and O-H$\cdots$O bonds  \cite{novak74:177}. Later, 
this study was extended to asymmetric HBs by Mikenda \cite{mikenda86:1}. 
For the case of N-H$\cdots$N bonds the correlation has been expressed previously 
by the function \cite{yan08:230}
\begin{equation}
\label{eq:fit}
f(r) = \omega_\infty/2\pi c \, {\rm erf}\left(\sum_{i=0}^2 a_i r^i\right),
\end{equation}
where $\omega_\infty$ is the  free NH stretching frequency in gas phase  
(for parameters see caption of Fig. \ref{fig:corr}). \Eq{eq:fit} can also 
be used for asymmetric HBs such as those of $\NHO$ type involving also HBs formed by amine groups (see Supplementary Information).
The advantage of such relation is obvious, it allows to assign stretching 
frequencies along a trajectory on the basis of  atomic coordinates only \cite{yan08:230}.
\subsection{IR Lineshape}
In the classical limit the IR lineshape can be calculated straightforwardly 
from the trajectory by means of the dipole-dipole correlation 
function \footnote{Note that the Gromacs/CPMD interface does not provide the 
capability to account for the QM charge density in the dipole calculation.}. 
Here we will use lineshape theory and account for the quantum nature of the 
high-frequency proton stretching vibrations. Specifically, we will calculate 
their gap correlation function, defined for some fundamental stretching frequency, 
$\omega_{10}(t)$, along a trajectory as follows:
\begin{equation}
\label{eq:corr}
C(t) = \left\langle \left(\omega_{10}(t)-\langle\omega_{10}\rangle\right)
\left(\omega_{10}(0)-\langle\omega_{10}\rangle\right)\right\rangle \, ,
\end{equation}
where $\langle\omega_{10}\rangle$ denotes the mean value along the trajectory. 
Assuming a second order cumulant expansion to hold the IR spectrum can be 
expressed via the lineshape function
\begin{equation}
\label{eq:gt}
g(t) \equiv \int^t_0d\tau\int^\tau_0d\tau^\prime\, C(\tau^\prime)\, ,
\end{equation}
as \cite{mukamel95}
\begin{equation}
\label{eq:irsp}
\sigma(\omega) = \frac{1}{\pi}\textrm{Re}\int^\infty_0 dt \,
	e^{i(\omega-\left\langle \omega_{10}\right\rangle) t - g(t)} \, .
\end{equation}
Of course, Eq.  (\ref{eq:corr}), is a classical quantity and does not satisfy 
detailed balance. There are various recipes to compensate for this deficiency 
by introducing quantum correction factors \cite{ramirez04:3973}. Recently, Skinner and coworker 
gave an interesting account on this issue which revealed that  for the 
exemplary case of HOD in D$_{2}$O quantum corrections have a only modest 
influence on the IR lineshape \cite{lawrence05:6720}. In order to take into 
account quantum effects we will fit the classical correlation function to 
the expression obtained within the  multimode oscillator model \cite{mukamel95}
\begin{eqnarray}
\label{eq:qcf} 
C(t) &=& \sum_jS(\omega_j)\omega^2_j\coth(\hbar\omega_j/2k_{\rm B}T)\cos(\omega_jt) \nonumber\\
&+& i \sum_jS(\omega_j)\omega^2_j\sin(\omega_jt) \, ,
\end{eqnarray}
where $S(\omega_j)$ is the dimensionless Huang-Rhys factor giving the coupling 
strength between the two-level system and the bath mode with frequency $\omega_j$.  
These two sets of parameters are used for fitting the real part of $C(t)$ obtained from 
the classical molecular dynamics which in turn gives the imaginary part as well.
%
\section{Results}
\subsection{Geometry-based Correlations}
In Fig. \ref{fig:dist}a and  \ref{fig:dist}b we depict the $\NN$ and the $\NO$ 
distance, respectively,  along the production part of the trajectory. 
In the course of the trajectory no H-atom transfer occurs. The average HB 
lengths are $L(\textrm{N}\cdots\textrm{N})=3.09$ \AA{}  and 
$L(\textrm{N}\cdots\textrm{O})=3.03$ \AA. Since geometry-based correlations are usually 
discussed assuming linear HBs we illustrate in panels (c) and (d) of 
Fig. \ref{fig:dist} the deviation of the H atoms from linear motion. 
The average deviations are 0.028 \AA{} for the $\NHN$ and 0.038 \AA{} for 
the $\NHO$ bond, that is, both cases can be considered to correspond to almost 
linear HBs. 

\begin{figure}[ht!]
\includegraphics[width=0.45\textwidth]{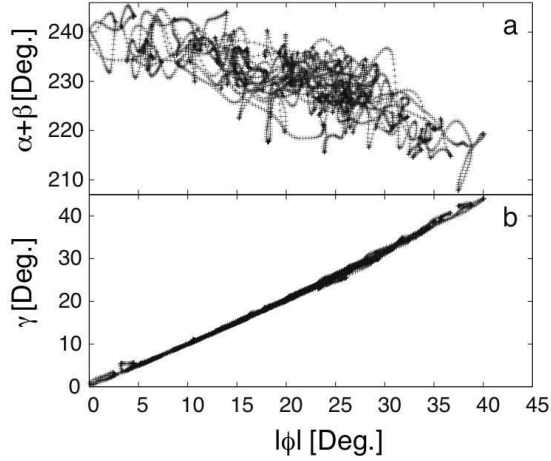}
\caption{Geometric correlations of the $\NHN$ and $\NHO$ HBs along the QM/MM 
trajectory between the absolute value of the dihedral angle $\phi$ and  
(a) the sum of the angles $\alpha$  and $\beta$ and 
(b) the angle $\gamma$  (for definitions see caption of Fig. \ref{fig:structure}).
}
\label{fig:geo}
\end{figure}

The planarity of the  gas phase A:U heterocylces is no longer present in 
solution. Focussing on the two HBs only, one can define the dihedral angle 
$\phi$ formed by the atoms N$_1$, N$_3$, O$_4$, and N$_6$ 
(see. Fig. \ref{fig:structure}). Its average value during the propagation 
interval is -19$^\circ$. Fig. \ref{fig:geo} shows how this deformation is 
related to other geometrical changes of the HBs. From panel (a) we see that 
nonplanarity comes along with a decrease of the angle $\alpha+\beta$, that is, 
with a compression of the HBs due to N$_3$H$_{\rm U}$ and/or C$_4$O$_4$ bending 
vibrations. Fig. \ref{fig:geo}b reveals an almost perfect linear correlation 
between the dihedral angle $\phi$ and the angle $\gamma$ describing 
the directions of the two HBs. 
\begin{figure}[t!]
\includegraphics[width=0.45\textwidth]{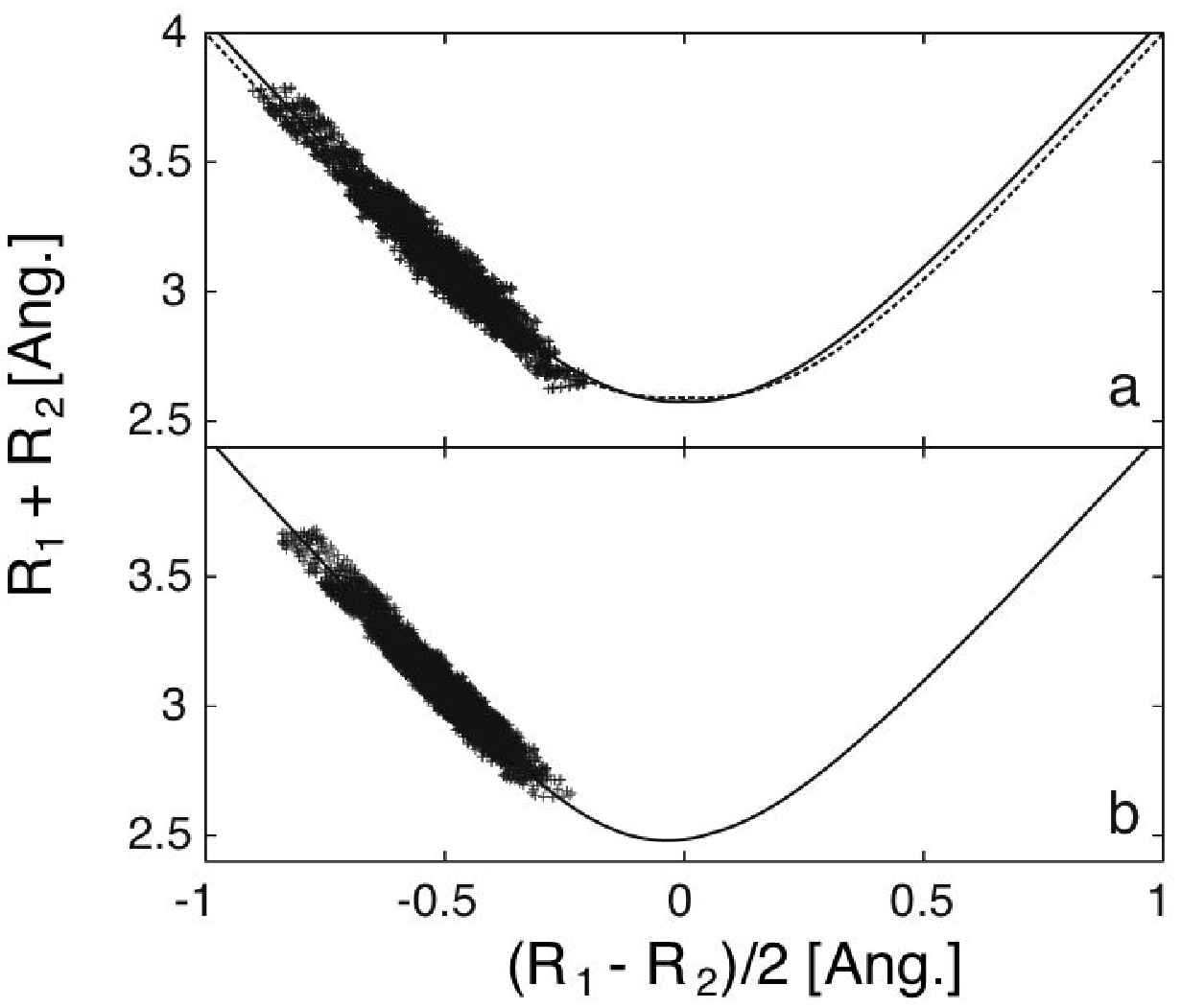}
\caption{Geometric correlations between HB length and H-atom position 
following from  a fit to \Eq{eq:bondordercor} (solid line) with the points 
sampled along the QM/MM trajectory. 
(a)  $\NHN$ ($1=2={\rm NH}$) with $R^{\rm eq}_1 = 1.025$ \AA, $b_1  = 0.361$ \AA,  $c  = 577.64$,  
and $d  = 0.30$ and (b) $\NHO$  ($1={\rm NH}$, $2={\rm OH}$) with 
$R^{\rm eq}_1  =  1.021$ \AA, $b_1$ =  0.407 \AA, $R^{\rm eq}_2$  =  1.029 \AA, 
$b_2$ =  0.232 \AA, $c$ =  400, $d$=  0.21. In panel (a) we also show the correlation 
curve from Limbach et al. (dashed line) \cite{pietrzak07:296}.
}
\label{fig:limbach}
\end{figure}
Besides the geometric correlation between two HBs, there also exists correlation for each HB separately.
Fig. \ref{fig:limbach} gives a test of the empirical correlation between N-H and HB 
length derived from \Eq{eq:bondordercor}. As anticipated from the robustness of 
the correlation reported before \cite{limbach04:115,limbach06:193,pietrzak07:296}, 
geometries along the trajectory are remarkably well described by this simple 
expression. This holds true, irrespective of the deviation from linearity and 
planarity mentioned above. Moreover, there seems to be little variance with 
respect to the class of molecules as shown by plotting the correlation curve with 
parameters obtained by Limbach et al. for intramolecular $\NHN$ HBs (dashed line in panel (a))  \cite{pietrzak07:296}.

\begin{figure}[t!]
\includegraphics[width=0.45\textwidth]{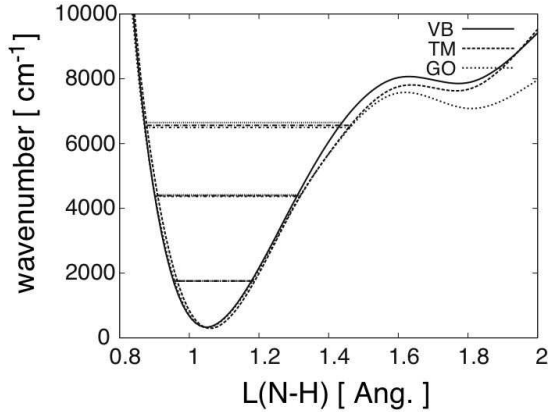}
\caption{Comparison of potential energy curves for the $\NHN$ HB as calculated from different
PPs by changing the N-H bond length (11 points in the given interval) 
with all other coordinates being fixed at the $t=0$ ps structure. 
The horizontal lines  correspond to the lowest vibrational states which have been 
obtained using the Fourier Grid Hamiltonian method \cite{marston89:3571}. The cut-off parameters are 
30 Ry for VB, 70 Ry for TM and 140 Ry for GO.
}
\label{fig:pes}
\end{figure}

In a next step we explore the correlation between HB bond lengths $\NN$ and $\NO$ 
and the fundamental transition frequencies of the NH stretching motion. 
A straightforward realization would consist  in the calculation of  NH 
stretching snapshot potentials in an otherwise frozen geometry for some 
representative points along the trajectory. Interpolating these data one 
will have at hand a distance-frequency correlation to be used for all points 
along the trajectory. Fig. \ref{fig:pes} depicts a representative potential for 
the $\NHN$ HB and its dependence on the form of the PP. First, 
we notice that up to the overtone excitation the potential is rather well 
described by an anharmonic oscillator and the second rather shallow minimum 
corresponding to an H transfer configuration is energetically not favorable. 
Second, apart from the shape of this shallow transfer minimum the three 
PPs give rather similar fundamental transition frequencies, which are 
2634 \cm for VB, 2647 \cm for TM and 2600 for GO.
Third, the fundamental transition frequency in the 2600 \cm range is considerable lower 
than the experimental value of 3185 \cm \cite{woutersen04:5381}. 

\begin{figure}[ht!]
\includegraphics[width=0.45\textwidth]{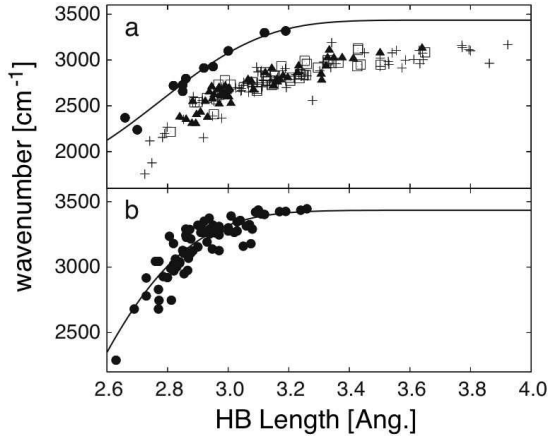}
\caption{
(a) Correlation between HB length $\NN$ and fundamental NH transition frequency (bullets: spectroscopic data for crystals containing intermolecular $\NHN$ 
HB \cite{novak74:177,yan08:230}; QM/MM on the fly  frequencies using the 
VB (crosses), TM (squares), and GO (triangles) PPs with cut-off of
30 Ry, 70 Ry, and 140 Ry, respectively. 
The solid line is a fit of experimental data according to \Eq{eq:fit} 
with $\omega_\infty/2\pi c = 3436$ \cm \cite{colarusso97:39}, $a_0 = 7.0911$,
$a_1 =-5.7941~\textrm{\AA}^{-1}$, and $a_2 = 1.2711~\textrm{\AA}^{-2}$.
(b) Correlation between HB length $\NO$ and fundamental N-H transition frequency 
(bullets: spectroscopic data for crystals containing intermolecular $\NHO$ HBs, 
solid line: fit according to \Eq{eq:fit} with
$\omega_\infty/2\pi c = 3434$ \cm \cite{colarusso97:39},
$a_0 =-0.5345$,
$a_1 =-0.8332~\textrm{\AA}^{-1}$, and
$a_2 = 0.5044~\textrm{\AA}^{-2}$).
}
\label{fig:corr}
\end{figure}

A more complete picture is obtained from  Fig. \ref{fig:corr}a where we show results from snapshot frequency calculations 
with the VB, TM and GO PPs 
sampled along the trajectory. They reproduce the expected qualitative dependence 
on the HB length, but are red-shifted by about 450 \cm  with respect to the 
empirical correlation curve obtained from crystal data. This 
holds irrespective of the PP and  cut-off parameter. 

The first question to be clarified concerns the experimental 
assignment in Ref. \cite{woutersen04:5381}. In fact  the similar case  of 
9-ethyladenine and cyclohexyluracil in chloroform has been studied previously 
and it was found that self-association of U:U dimers occurs which would have an 
NH stretching absorption in the 3000-3200 \cm range as well. However, 
the association constant for U:U dimers is 15 times smaller than for A:U ones, 
so that the contribution from U:U dimers should be negligible  \cite{kyogoku67:969}. 
A similar conclusion can be reached based on the work of Miller et al. 
who demonstrated an 1:1 stoichiometry for similar purine and pyrimidine base 
analogues in CDCl$_3$ by monitoring the IR absorption changes around 
3210 \cm as a function of the uracil mole fraction \cite{miller67:345}. 

Giving this evidence for the correct experimental assignment, we compare the present 
findings with recent gas phase quantum chemical studies of the thymine NH 
stretching vibration in the  adenine:thymine base pair \cite{krishnan07:132}. 
In harmonic approximation the DFT/B3LYP (6-31++G(d,p)) level of theory yields 
a frequency of 2981 \cm. Accounting for anharmonicity within a three-dimensional 
model which includes all H-bonded stretching vibrations lowers this value to 
2608 \cm. In order to rule out that this is a result of too soft DFT potential 
energy surfaces, MP2 corrections have been added to the diagonal anharmonicity. 
However, this causes an increase of the thymine stretching frequency to 
2688 \cm only. Very recently, Yagi and coworkers have started to investigate 
this vibration using their VMP2-(4) method \cite{yagi07:034111} on the basis of 
a B3LYP/6-31G++(d,p)  full-dimensional quartic force field potential. 
They obtained a preliminary value as low as 2488 \cm \cite{yagi08:}. 
At this level of theory the  $\NN$ distance is 2.88 \AA{}, which is smaller than 
the average value obtained from the QM/MM trajectory. Therefore, the lower value 
of the NH stretching frequency in the gas phase is consistent with the stronger 
HB as compared to the solution phase. However, giving the empirical correlations discussed in the following the absolute
value of the calculated frequencies appears to be considerably too small.

\begin{figure}[ht!]
\includegraphics[width=0.45\textwidth]{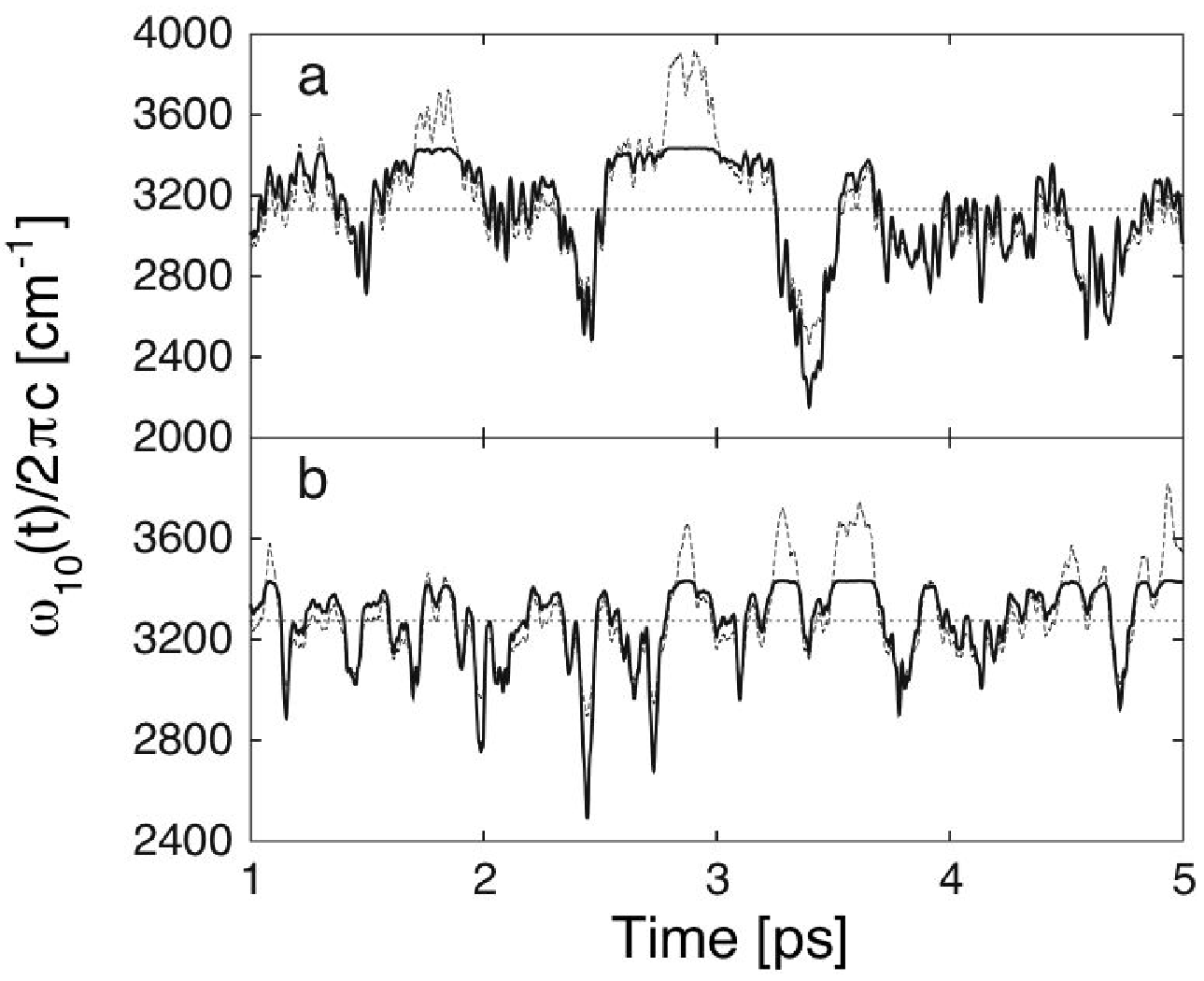}
\caption{Time-dependent fundamental transition frequencies for the $\NHN$ (a) 
and the $\NHO$ (b)  HB as obtained using the empirical correlations displayed in
Fig. \ref{fig:corr}. The dotted lines are the averaged values 
$\langle \omega_{10}\rangle$ which are equal to 3132 \cm in (a) and 3275 \cm in (b). 
The dashed lines are the $\NN$ and $\NO$ distances scaled and shifted 
such as to reveal the correlation with the time-dependent frequencies. 
}
\label{fig:omegat}
\end{figure}

To cope with this problem we decided to resort to an empirical mapping of 
HB distances to the stretching frequencies. In Ref. \cite{yan08:230} we have 
shown that this correlation plotted in Fig. \ref{fig:corr}a gives an average 
frequency of 3227 \cm for the VB PP with a cut-off of 30 Ry and using a 6.2 ps trajectory. Furthermore, the lineshape 
derived from the frequency fluctuations has been in good agreement with the 
experiment as well. This supports the assumption that despite the discrepancy 
in the absolute value of the stretching frequency the current QM/MM protocol 
provides a reasonable description of the fluctuations influencing the HB. 
We have extended this approach to the case of the $\NHO$ bond; 
the frequency-HB length correlation plot derived from available crystal data 
is shown in Fig. \ref{fig:corr}b. Notice that the data used included HBs 
involving amine groups, that is, the fact that the considered vibrations 
is actually of symmetric NH$_2$ character is taken care of by the empirical relation. 
The resulting time-dependent frequencies for the two HBs are given in  Fig. \ref{fig:omegat}.
\subsection{IR Lineshape Analysis}
\begin{figure}[t!]
\includegraphics[width=0.45\textwidth]{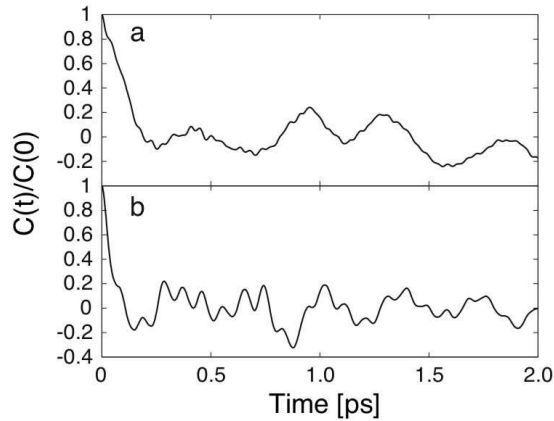}
\caption{
Frequency correlation function, \Eq{eq:corr} for the NH stretching vibrations of (a) the 
 $\NHN$ and (b) the  $\NHO$ HB.}
\label{fig:ct}
\end{figure}

\begin{figure}[ht!]
\includegraphics[width=0.45\textwidth]{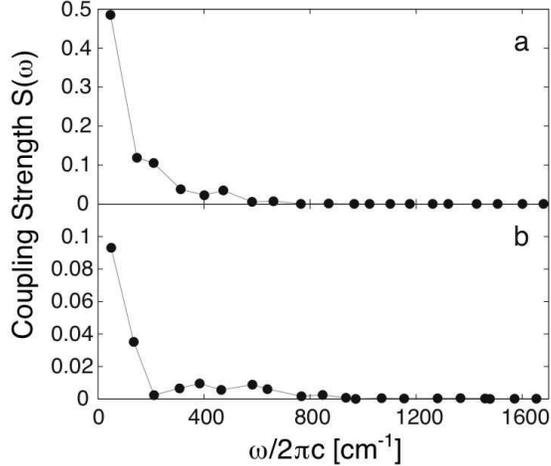}
\caption{
Coupling strength between the NH vibration and the bath
as function of the bath frequency. 
(a) results for $\NHN$ HB; (b) results for $\NHO$ HB.
Points are fitted results and line is guide for eye.
}
\label{fig:coupl}
\end{figure}

The correlation function $C(t)$ has been calculated for both HBs for the gap fluctuations 
shown in  Fig. \ref{fig:omegat}. The results  given in 
Fig. \ref{fig:ct} reveal a rapid initial decay followed by pronounced 
oscillations lasting up to the 2 ps which can be covered by the available 
trajectory (for a convergence study, see also Ref. \cite{yan08:230}). 
In order to gain 
further insight into the distribution of modes coupled to the considered 
transitions, we have fitted $C(t)$ according to the Brownian oscillator model, 
\Eq{eq:qcf}. The chosen frequency range of 0-1770 \cm covers the low-frequency 
part of the calculated gas phase harmonic IR spectrum of the A:U pair. It is discretized into 20 modes whose frequencies 
and coupling strengths are used as fitting parameters. The distribution of coupling 
strengths $S(\omega)$ is shown in Fig. \ref{fig:coupl}. First, we notice that 
both distributions are dominated by low-frequency vibrations. 
Second, in both cases we observe two smaller peaks which are around 200 and 500 \cm 
for $\NHN$ as well as around  350 and 550 \cm for $\NHO$. It is tempting 
to assign these peaks to intermolecular vibrational modes. Indeed gas phase 
harmonic analysis of the A:U pair gives several intermolecular modes modulating 
the $\NN$ and $\NO$ distances in region of 150-400 \cm. Modes in the range of 
400-600 \cm are mostly modulating the $\NO$ distance consistent with Fig. \ref{fig:coupl}. 

\begin{figure}[ht!]
\includegraphics[width=0.45\textwidth]{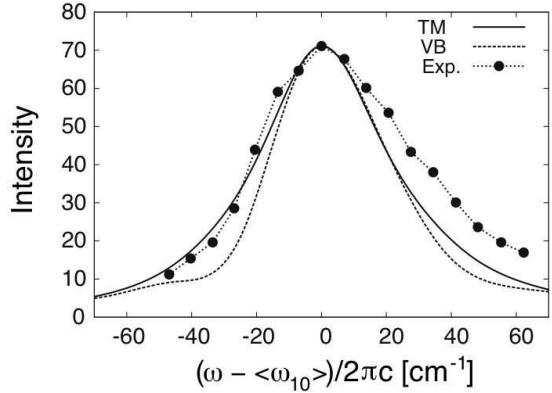}
\caption{
IR lineshape of NH stretching compared with experimental 
data~\cite{woutersen04:5381} (line for experimental result
is drawn as guide for the eye only). Note that we have not included the shift of the transition frequency due to the imaginary
part of the correlation function. It amounts to -3 \cm for the TM and -16 \cm for the VB case.
}
\label{fig:spec}
\end{figure}

From the correlation functions one obtains the IR lineshape which is exemplarily 
shown for the $\NHN$ case in Fig. \ref{fig:spec}.
First we notice that the overall agreement with the experimental data is 
fairly good given the simple empirical model for assigning transition frequencies. 
In particular the asymmetry of the line is well reproduced. 
We also show the previous result obtained using the VB 
PP \cite{yan08:230} which merely gives a slightly lower value for 
the FHWM. 
\begin{table*}[t]
\caption{\label{tab:comp}Comparison of HB geometry and lineposition and lineshape 
for different PPs and cut-off parameters. 
Labels $\textrm{L}_{\textrm{N}\cdots\textrm{N/O}}$, $\nu_{\textrm{NH}\cdots\textrm{N/O}}$
and $\Gamma_{\textrm{N}\cdots\textrm{N/O}}$
stand for bond length, fundamental transition frequency and the FHWM of the IR lineshape 
for $\textrm{NH}\cdots\textrm{N/O}$ HB,  respectively. Gap refers to the distance
between the fundamental transition frequency of $\NHN$ and $\NHO$ HBs and 
$\Gamma^{\rm AI}_{\textrm{N}\cdots\textrm{N}}$ to the FHWM of the $\NHN$ HB 
based on the stretching-distance correlation from snapshot potentials. The TM (70 Ry) setup 
has been used in Fig. \ref{fig:spec}. Note that the frequencies do not contain the shift due to the imaginary part of the correlation function which is smaller than 5 \cm in all cases. Further notice that the VB 30 Ry results are slightly different from Ref. \cite{yan08:230} where a  6.2 ps trajectory had been used.
}
\begin{ruledtabular}
\begin{tabular}{cc|cc|ccc|ccc}
PP & cut-off & $\textrm{L}_{\textrm{N}\cdots\textrm{N}}$ & $\textrm{L}_{\textrm{N}\cdots\textrm{O}}$ & $\nu_{\textrm{NH}\cdots\textrm{N}}$ & $\nu_{\textrm{NH}\cdots\textrm{O}}$  & gap & $\Gamma_{\textrm{N}\cdots\textrm{N}}$ & $\Gamma_{\textrm{N}\cdots\textrm{O}}$ & $\Gamma^{\rm AI}_{\textrm{N}\cdots\textrm{N}}$ \\
   & (Ry)   & (\AA)        & (\AA)        & (\cm)           & (\cm)  & (\cm) & (\cm)    & (\cm)  & (\cm) \\\hline
  VB&   30&   3.24&   3.11&   3235&   3306&     71&     37&     9&     47  \\
  VB&   40&   3.09&   3.14&   3127&   3306&    179&     41&     13&     37  \\
\textbf{TM}&   \textbf{70}&   \textbf{3.09}&   \textbf{3.03}&   \textbf{3132}&   \textbf{3275}&    \textbf{143}&     \textbf{45}&     \textbf{10}&    \textbf{41}  \\
   TM&  100&   3.10&   3.09&   3122&   3276&    154&     54&      8&     55  \\
   GO&  100&   3.10&   2.99&   3183&   3250&     67&     23&      9&     19  \\
   GO&  140&   3.09&   3.05&   3143&   3293&    150&     15&      5&     13  \\
   \hline\hline
 Exp. &     - &    - & - &       3185 & 3315 & 130 & 53 & 41  & 53  \\
\end{tabular}
\end{ruledtabular}
\end{table*}

A detailed comparison of lineshape parameters obtained by using different 
PPs and cut-off parameters is presented in Tab. \ref{tab:comp}. 
If we take the average transition frequencies and the corresponding energy gap 
between the two stretching vibrations as the measure for the performance of the method, 
we find that the TM PP with a cut-off of 70 Ry gives the best results. 
Here the deviations from experimental values are below 2 \% for both frequencies.
Of course, this result is not general in the sense that it is based on the 
proposed simulation protocol which includes the empirical mapping of distances 
to frequencies. Therefore using the larger cut-off of 100 Ry does not necessarily 
give an improvement. Another point to mention is that Tab. \ref{tab:comp} also 
shows that the correlation between average HB distance and average stretching 
frequency is not linear and in particular the fluctuations are sensitive to 
the used CPMD setup. The latter are also responsible for the linewidths which 
are compared in the right part of Tab. \ref{tab:comp}. Also for this quantity 
the TM PP (70 Ry) performs reasonably well for the case of the 
$\nu_{\NHN}$ transition. The FWHM of the 
$\nu_{\NHO}$ transition is, however, considerably 
underestimated in all cases.
Finally, we give the FWHM obtained by using the gap correlation function derived from 
the snapshot potentials. They are remarkably close the values calculated via 
the empirical correlation. This leads us to conclude that the quantitative 
difference between the two approaches is indeed merely a constant shift of the 
transition frequencies (cf. Fig. \ref{fig:corr}).
%
\section{Summary}
%
We have presented a QM/MM approach to the simulation of the intermolecular HB 
dynamics in solvated A:U pairs. On average the two HBs are almost linear and follow 
the empirical geometric relation of Limbach and coworkers between the HB lengths 
and the proton position. Further it was found that nonplanarity of the HBs is 
closely related to the compression due to bending motions of the C=O and 
NH groups. The trajectory mean of the dihedral angle formed by HB donors and acceptors 
is predicted as -19$^\circ$. Exploring the empirical correlation between HB lengths and 
transition frequencies of the H-bonded protons we were able to simulate the 
IR spectrum in the region of the uracil NH and the adenine symmetric NH$_2$ 
stretching fundamental transition. Attempts to use transition frequencies 
derived from snapshot potentials gave considerably red-shifted transitions 
independent on the used pseudopotential and cut-off parameters. The difference 
between empirical and snapshot-derived frequency-distance correlation curves 
amounts to a shift of 450 \cm approximately independent on the HB distance. 
For the used empirical mapping the Troullier-Martins pseudopotential with 
a cut-off of 70 Ry gives the best agreement for the two fundamental transition. 
A multimode Brownian oscillator analysis of the gap correlation function yielded 
a spectrum of coupling strengths which is dominated by low-frequency vibrations, 
but contains contributions presumably from modes which modulate the two HBs in 
the range between 200-700 \cm. While the width of the  
$\nu_{\textrm{NH}\cdots\textrm{N}}$ transition could be fairly well reproduced, 
the width of the $\nu_{\textrm{NH}\cdots\textrm{O}}$ transition is underestimated. 
However, one should be aware that our simulation does neither include 
contributions from other high-frequency modes nor combination bands which could 
give rise to an additional broadening of the experimental spectra. 
%
\begin{acknowledgments}
We are indebted to Dr. S. Woutersen (Amsterdam) for the helpful discussion 
concerning the NH assignment and Dr. Yagi (Tokyo) for sharing his preliminary 
results on the anharmonic vibrational calculations. This work has been 
financially supported by the Deutsche Forschungsgemeinschaft (project Ku952/5-1).
\end{acknowledgments}

\end{document}